\documentclass[a4paper,12pt]{article}
\pdfoutput=1
\usepackage{graphicx, rotating}
\usepackage{hyperref}
\usepackage{slashed}
\usepackage{ifpdf}

\ifx\pdfoutput\undefined
\usepackage[dvips,bookmarks=false]{hyperref}	
\else
\usepackage{hyperref}	
\fi
\hypersetup{colorlinks,bookmarksopen,bookmarksnumbered,citecolor=verdes,
linkcolor=blus,pdfstartview=FitH,urlcolor=rossos}
\def\hhref#1{\href{http://arxiv.org/abs/#1}{#1}} 

\usepackage{amsmath}
\usepackage{slashed}

\newcommand{\beq}{\begin{equation}}
\newcommand{\eeq}{\end{equation}}
\newcommand{\fig}[1]{~\ref{fig:#1}}

\newcount\Mac  \Mac=1  
\newcommand{\ifMac}[2]{\ifnum\Mac=1 #1 \else #2 \fi}
\def\putps(#1,#2)(#3,#4)#5#6{\ifnum\Mac=1 \put(#1,#2){\special{picture #5}}
\else  \put(#3,#4){\includegraphics{#6}} \fi}

\newcommand{\One}{\hbox{1\kern-.24em I}}

\newcommand{\pb}{\,{\rm pb}}
\newcommand{\GeV}{\,{\rm GeV}}

\newcommand{\MeV}{\,{\rm MeV}}

\newcommand{\eq}[1]{~{\rm(\ref{eq:#1})}}

\newcommand{\lascia}[1]{}
\makeatletter
\def\art{\@ifnextchar[{\eart}{\oart}}
\def\eart[#1]#2#3#4#5#6{{\rm #2}, {#3 #4} {\rm (#6) #5} [arXiv:{\hhref{#1}}]}
\def\hepart[#1]#2{{\rm #2, arXiv:\hhref{#1}}}
\newcommand{\oart}[5]{{\rm #1}, {#2 #3} {\rm (#5) #4}}

\newcounter{alphaequation}[equation]
\def\thealphaequation{\theequation\hbox to
0.6em{\hfil\alph{alphaequation}\hfil}}
\def\eqnsystem#1{
\def\@eqnnum{{\rm (\thealphaequation)}}
\def\@@eqncr{\let\@tempa\relax \ifcase\@eqcnt \def\@tempa{& & &} \or
  \def\@tempa{& &}\or \def\@tempa{&}\fi\@tempa
  \if@eqnsw\@eqnnum\refstepcounter{alphaequation}\fi
\global\@eqnswtrue\global\@eqcnt=0\cr}
\refstepcounter{equation} \let\@currentlabel\theequation \def\@tempb{#1}
\ifx\@tempb\empty\else\label{#1}\fi
\refstepcounter{alphaequation}
\let\@currentlabel\thealphaequation
\global\@eqnswtrue\global\@eqcnt=0 \tabskip\@centering\let\\=\@eqncr
$$\halign to \displaywidth\bgroup \@eqnsel\hskip\@centering
$\displaystyle\tabskip\z@{##}$&\global\@eqcnt\@ne
\hskip2\arraycolsep\hfil${##}$\hfil& \global\@eqcnt\tw@\hskip2\arraycolsep
$\displaystyle\tabskip\z@{##}$\hfil
\tabskip\@centering&\llap{##}\tabskip\z@\cr}

\def\endeqnsystem{\@@eqncr\egroup$$\global\@ignoretrue} \makeatother

\def\circa#1{\,\raise.3ex\hbox{$#1$\kern-.75em\lower1ex\hbox{$\sim$}}\,}

\usepackage{multicol}
\usepackage{color}
\definecolor{rosso}{cmyk}{0,1,1,0.4}
\definecolor{rossos}{cmyk}{0,1,1,0.55}
\definecolor{rossoc}{cmyk}{0,1,1,0.2}
\definecolor{blu}{cmyk}{1,1,0,0.3}
\definecolor{blus}{cmyk}{1,1,0,0.6}
\definecolor{bluc}{cmyk}{1,1,0,0.1}
\definecolor{verde}{cmyk}{0.92,0,0.59,0.25}
\definecolor{verdec}{cmyk}{0.92,0,0.59,0.15}
\definecolor{verdes}{cmyk}{0.92,0,0.59,0.4}
\definecolor{grigio}{cmyk}{0,0,0,0.07}
\definecolor{rosa}{cmyk}{0,0.1,0.1,0.02}
\definecolor{rosino}{cmyk}{0,0.05,0.05,0.02}
\definecolor{rosas}{cmyk}{0,0.3,0.25,0.05}
\definecolor{celeste}{cmyk}{0.1,0,0,0.02}
\definecolor{giallino}{cmyk}{0,0,0.4,0.02}
\definecolor{rosso}{cmyk}{0,1,1,0.4}
\definecolor{rossos}{cmyk}{0,1,1,0.55}
\definecolor{rossoc}{cmyk}{0,1,1,0.2}
\definecolor{blu}{cmyk}{1,1,0,0.3}
\definecolor{bluc}{cmyk}{1,1,0,0.1}
\definecolor{blucc}{cmyk}{0.7,0.5,0,0}
\definecolor{viola}{cmyk}{0,1,0,0.6}
\definecolor{viola2}{cmyk}{0,1,0.2,0.6}
\definecolor{verde}{cmyk}{0.92,0,0.59,0.25}
\definecolor{verdec}{cmyk}{0.92,0,0.59,0.15}
\definecolor{verdes}{cmyk}{0.92,0,0.59,0.4}
\definecolor{verdino}{cmyk}{0.12,0,0.09,0.05}
\definecolor{giallo}{cmyk}{0,0,1,0}
\definecolor{gialloverde}{cmyk}{0.44,0,0.74,0}

\font\tenrsfs=rsfs10 at 12pt

\font\sevenrsfs=rsfs7
\font\fiversfs=rsfs5
\newfam\rsfsfam
\textfont\rsfsfam=\tenrsfs
\scriptfont\rsfsfam=\sevenrsfs
\scriptscriptfont\rsfsfam=\fiversfs
\def\mathscr#1{{\fam\rsfsfam\relax#1}}

\def\eq#1{eq.~(\ref{#1})}
\def\beq{\begin{equation}}
\def\eeq{\end{equation}}
\def\bea{\begin{eqnarray}}
\def\eea{\end{eqnarray}}

\begin{document}\hfill
 \centerline{CERN-PH-TH/2012-066}

\color{black}
\vspace{1cm}
\begin{center}
{\LARGE\bf\color{black}Reconstructing Higgs boson properties\\[3mm] from the LHC and Tevatron data}\\
\bigskip\color{black}\vspace{0.6cm}{
{\large\bf Pier Paolo Giardino$^{a}$, Kristjan Kannike$^{{b,c}}$, \\[3mm]
Martti Raidal$^{c,d,e}$
 {\rm and} Alessandro Strumia$^{a,c}$}
} \\[7mm]
{\it (a) Dipartimento di Fisica dell'Universit{\`a} di Pisa and INFN, Italy}\\[1mm]
{\it  (b) Scuola Normale Superiore and INFN, Piazza dei Cavalieri 7, 56126 Pisa, Italy}\\[1mm]
{\it  (c) National Institute of Chemical Physics and Biophysics, Ravala 10, Tallinn, Estonia}\\[1mm]
 {\it (d) CERN, Theory Division, CH-1211 Geneva 23, Switzerland}\\[1mm]
{\it  (e) Institute of Physics, University of Tartu, Estonia}\\[3mm]
\end{center}
\bigskip
\bigskip
\bigskip
\vspace{1cm}

\centerline{\large\bf\color{blus} Abstract}

\begin{quote}
We perform a phenomenological fit to all ATLAS, CMS, CDF and D0 Higgs boson data available after Moriond 2012.
We allow all Higgs boson branching fractions, its  couplings to standard model particles, as well as to an hypothetical invisible sector to vary freely,
and determine their current favourite values.  The standard model Higgs boson with a mass 125~GeV correctly predicts the average observed rate
and provides an acceptable global fit to data. However,  better fits are obtained by non-standard scenarios that 
reproduce anomalies in the present data (more $\gamma\gamma$ and less $WW$ signals than expected)
such as modified rates of loop processes or partial fermiophobia. We find that present data disfavours Higgs boson invisible decays. 
We consider implications  for the standard model,  for supersymmetric and fermiophobic Higgs bosons,  
 for dark matter models, for warped extra-dimensions.
\end{quote}

\newpage


\section{Introduction}
Identifying the mechanism of electroweak symmetry breaking is the main goal of the Large Hadron Collider (LHC).
In the standard model (SM) the electroweak symmetry is broken due to the existence of an elementary scalar particle ---
the Higgs boson \cite{Englert:1964et,Higgs:1964ia,Higgs:1964pj,Guralnik:1964eu,rev}. 
Based on data collected in 2011, both the ATLAS and CMS experiments at the LHC  published results of their searches 
for the SM-like Higgs boson that, yet  inconclusively, support its existence with a mass $m_h\approx 125$~GeV~\cite{mhexp,Acomb,Ccomb}.   
Those results have been recently updated at the Moriond 2012 conference, where all the Tevatron 
and LHC collaborations presented their updated Higgs boson searches as well as some  new results. 
The combined Tevatron analysis of all collected data confirms the LHC excess around 125~GeV in the $h\to b\bar b$ channel at $2.6\sigma$ level;
CMS presented an improved $\gamma\gamma$ analysis; ATLAS presented new $WW^*$, $b\bar b$ and $\tau\bar\tau$ searches with full 2011 luminosity. 
Furthermore, both the ATLAS and CMS experiments showed  results of searches for a fermiophobic (FP) Higgs boson in 
the $h\to \gamma \gamma$ channel that both show a positive hint  around 125~GeV with local significances about $3\sigma$. 
This excess is consistent with the total inclusive $\gamma\gamma$ rate observed by the LHC~\cite{Gabrielli:2012yz}.

Accidentally, $m_h\approx 125$~GeV is a particularly fortunate value for the LHC,
because, according to the SM predictions, various Higgs boson search channels are measurable. 
Those  arise from a combination of 
SM Higgs boson branching fractions~\cite{BR}
\beq \begin{array}{lll}
\hbox{BR}(h\to b\bar b) =58\%,\qquad&
\hbox{BR}(h\to WW^*)=21.6\%, \qquad&
\hbox{BR}(h\to\tau^+\tau^-)=6.4\%, \\
\hbox{BR}(h\to ZZ^*) =2.7\%,\qquad&
\hbox{BR}(h\to gg)=8.5\% ,&
\hbox{BR}(h\to \gamma\gamma)=0.22\%, \\
\hbox{BR}(h\to c \bar{c})=2.7\%
\end{array}\eeq
and production mechanisms with cross sections~\cite{crosssections}
\beq \begin{array}{ll}
\sigma(pp\to h) =(15.3\pm2.6)\pb,\qquad&
\sigma(pp\to jjh)=1.2\pb,\\
\sigma(pp\to Wh)=0.57\pb,\qquad&
\sigma(pp\to Zh) = 0.32\pb ,\end{array}\eeq
named gluon-gluon fusion ($gg\to h$), vector-boson fusion (VBF) and 
associated production with $W$ and $Z$ bosons (Vh).
Because different search categories  are sensitive to different Higgs boson couplings, 
the LHC can study the properties of a Higgs boson with $m_h\approx 125\GeV$ and
test if it follows the SM predictions or is affected by new physics.



With the presently collected statistics none of the search channels alone is sensitive to the SM Higgs boson nor
are the combined results of Tevatron, ATLAS and CMS statistically conclusive. 
Therefore one expects large statistical fluctuations of the expected signal in all the search channels. 
Indeed, all measured LHC $\gamma\gamma$ rates, dominated by the new  results in the VBF category, 
have central values  above the SM prediction while all the $WW^*$ rates have central values consistently below the SM prediction.
On the one hand, those anomalies may be statistical fluctuations. On the other hand, they may signal  
new physics beyond the SM.
From a theoretical point of view, 
reconstructing the Higgs boson properties is an important way to address 
the main issue that LHC can clarify:
is there a natural reason behind the the smallness of the weak scale, $m_h \ll M_{\rm Pl}$?
Indeed, if the weak scale is  naturally small, one expects that
the new physics that cuts off the top loop contribution to $m_h^2$
(such as light stops at the weak scale in supersymmetric models)
also affects the $gg\to h$ and $h\to \gamma\gamma$ rates.
Therefore a global study of all the Higgs boson collider data obtained so far is necessary 
to test the SM and to discriminate between different new physics scenarios in the Higgs sector.


In this work we  study the collider data collected so far in Tevatron and the LHC in order to derive Higgs boson  properties.
The study of the Higgs boson fit was pioneered in Ref.~\cite{hfit}, while recent fits of new LHC data were published in Refs.~\cite{Contino,Falkowski,Grojean}.
We improve  on previous fits by including the new data presented in the Moriond 2012 conference, 
and by performing  more general fits that cover a wider spectrum of new physics models.
To achieve this goal we allow all the Higgs boson couplings 
to deviate independently  from their SM values. 
We also allow for an additional Higgs boson invisible width, possibly due to decays into the dark matter.
Anomalous features are dominated by the new results presented in Moriond 2012, disfavouring the SM compared to the previous fits
and motivating new physics scenarios. We discuss implications of our results in the context of different models. 
More LHC data is needed to discriminate between those scenarios.

The paper is organised as follows. 
In section~\ref{stat} we describe the  existing experimental results and the statistical procedure we adopt.
In section~\ref{res} we perform  the fits to data. In section~\ref{dis} we  discuss implications of our results on different models.
We conclude in section~\ref{concl}.

\section{Data and statistical analysis}
\label{stat}

The experimental collaborations measure rates of Higgs boson signals $R$.
Their results could be fully encoded in a likelihood ${\cal L}(R,m_h)$, 
but only a limited amount of information is reported by the experiments.
All collaborations report the upper bounds on rates at 95\% C.L., $R_{\rm observed}$, 
and the expected upper bound at $95\%$ C.L. in absence of a Higgs boson signal, $R_{\rm expected}$,
as function of the Higgs boson mass $m_h$. Given that information, our statistical analyses follows the one outlined in Ref.~\cite{Farina:2011bh}. 
Assuming that the $\chi^2 = -2 \ln {\cal L}$ has a Gaussian form in $R,$
\beq 
\chi^2 = (R-\mu)^2/\sigma^2,
\eeq
these two experimental informations allow one to extract the mean $\mu$ and the standard deviation $\sigma$,
\beq 
\mu = R_{\rm observed} - R_{\rm expected},\qquad
\sigma = \frac{R_{\rm expected}}{1.96},
\eeq
where $1.96$ arises because $95\%$ confidence level corresponds to about 2 standard deviations.
The Gaussian approximation by construction agrees with the full result at this value of $R$, but 
away from it 
the approximation may be not accurate for channels that presently have a low number of events (such as $h\to ZZ^*\to 4\ell$).
We are aware of this fact, but at the moment it is difficult to do better using the available data.
We verified that our procedure gives similar results as  the refined procedure in~\cite{Contino} and that
our procedure agrees better with present values of $\mu\pm\sigma$, when reported by experiments at $m_h=125\GeV$.

We also neglect correlations of uncertainties among different measurements (e.g.\
uncertainties on luminosity and on the SM prediction, at the $\pm15\%$ level and therefore 
subdominant with respect to present experimental
uncertainties) and approximate the full $\chi^2$ with
\beq
\chi^2 = \sum_i \frac{(R_i - \mu_i)^2}{\sigma_i^2},
\eeq
where the sum runs over all measured Higgs boson rates $i$.
In the present stage of experimental accuracy such a simplified statistical framework captures the main
features in data and allows us to study general properties of the data, that is the purpose of this work.

We consider all available Higgs boson data reported at the Moriond 2012 conference and before:
\begin{enumerate}
\item The $p\bar p\to Vh \to V b\bar b $ rate  measured by CDF, D0~\cite{bbTeVatron} and the related 
$pp\to Vh \to V b\bar b $ rate measured by CMS and ATLAS~\cite{bbCMS}.

\item The $pp\to jjh\to jjWW $ rate measured by CMS~\cite{jjh}.

\item The $h\to WW\to 2\ell2\nu$ rates measured by CMS and ATLAS~\cite{WW}.

\item The $h\to ZZ\to 4\ell$ rates measured by CMS and ATLAS~\cite{ZZ}. 
 
 \item The $h\to\gamma\gamma$ rates measured by ATLAS and CMS~\cite{gammagamma}, and CDF, D0 \cite{gammagammaTevatron}.
 
 \item  In the context of fermiophobic Higgs boson searches, CMS  measured the
 $pp\to jjh\to jj\gamma\gamma $ rate~\cite{gammagammajj} 
where the $jj$ tagging is added to select Higgs boson produced via the VBF process.
Indeed,  the cuts performed by CMS ($m_{jj}>350\GeV$, $p_{Tj_1}>30\GeV$, $p_{Tj_2}>20\GeV$)
significantly reduce the $gg\to h$
 contribution, such that we estimate that
reinterpreting this experimental result in a general context, it is roughly a measurement of
\beq[0.033 \sigma(pp \to h) +  \sigma(pp\to jjh)] \times\hbox{BR}(h\to\gamma\gamma)\eeq
with the result
\beq 
\frac{\hbox{observed rate}}{\hbox{SM rate}} = 3.3\pm 1.1\quad\hbox{for $m_h=125\GeV$}.
\eeq



\item  In the context of fermiophobic Higgs boson searches, ATLAS
measured the $pp\to hX\to \gamma\gamma X$ rate with a high cut $p_{Th} >40\GeV$
on the Higgs boson transverse momentum \cite{ATLAS_FB}
(we are oversimplifying by omitting several secondary issues).
This cut allows to suppress the $gg\to h$ production process, while keeping most of the signal in the VBF and associate production
mechanisms.
To see how much $gg\to h$ is suppressed we allowed for additional QCD jets performing
simulations with the {\sc Pythia}~\cite{Pythia} and {\sc MadGraph}~\cite{MadGraph} codes.
We find that this experimental result can be re-interpreted in a general context as a measurement of
\beq[0.3 \sigma(pp \to h) +  \sigma(pp\to Wh,Zh,jjh)] \times\hbox{BR}(h\to\gamma\gamma),\eeq
with the result
\beq \frac{\hbox{observed rate}}{\hbox{SM rate}} = 3.3\pm 1.1\quad\hbox{for $m_h=125\GeV$}.\eeq

 \item The $h\to\tau\tau$ rate as measured by CMS and ATLAS~\cite{tautau}.

\end{enumerate}

\begin{figure}
$$\raisebox{-0.2cm}{\includegraphics[height=6cm]{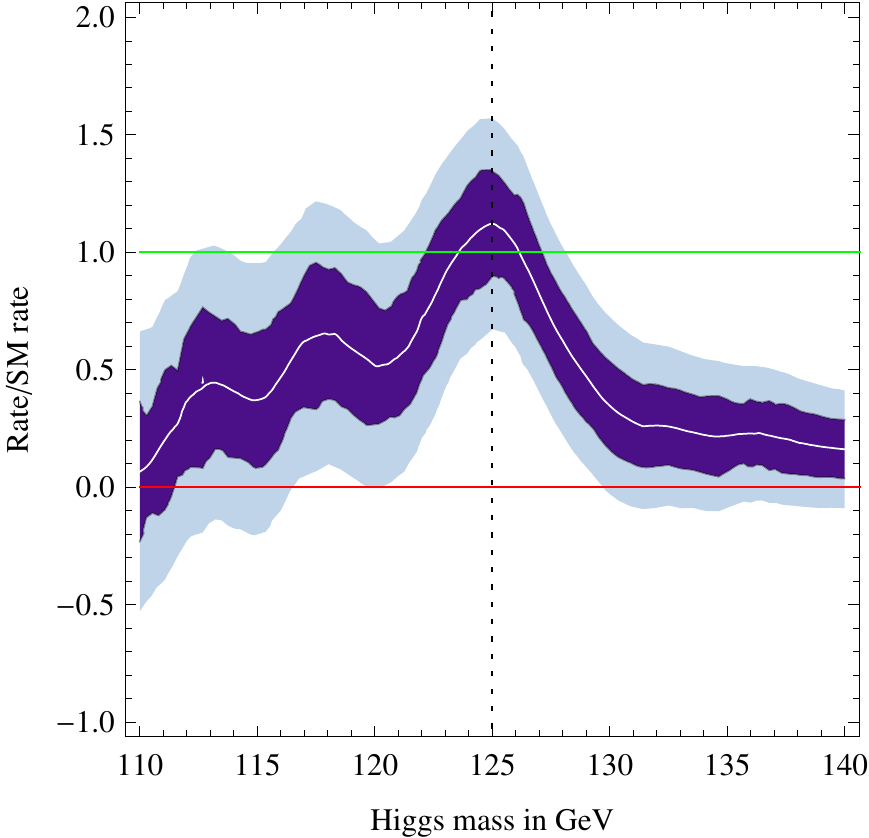}}\qquad \includegraphics[height=6.3cm]{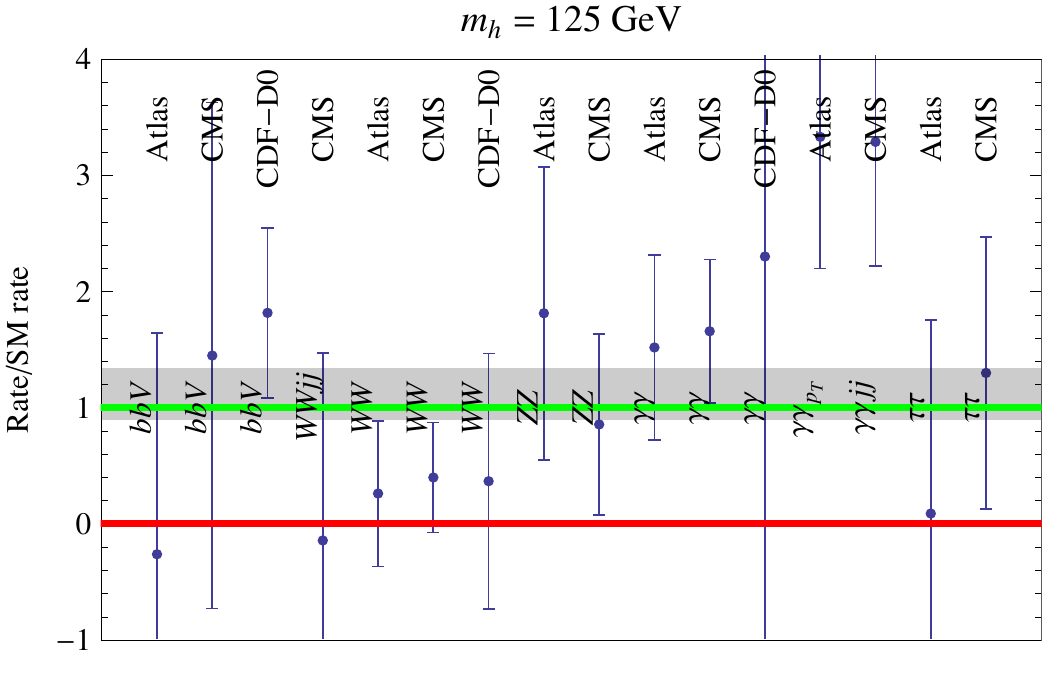}$$ 
\caption{\em {\bf Left}: The Higgs boson rate favoured at $1\sigma$ (dark blue) and $2\sigma$ (light blue) in a global SM fit as function of the Higgs boson mass.
{\bf Right}: assuming $m_h=125\GeV,$ we show the
measured Higgs boson rates at ATLAS, CMS, CDF, D0 and their average (horizontal gray band at $\pm1\sigma$).
Here 0 (red line) corresponds to no Higgs boson, 1 (green line) to the SM Higgs boson.
\label{fig:data}}
\end{figure}
In the left panel of Fig.\fig{data} we show our approximated combination of all Higgs boson data.
Higgs boson masses  around $125$~GeV are favoured by the rate, and some $ZZ$ and $\gamma\gamma$ events
(which have little statistical power in fixing the rates but large resolution in $m_h$)
favor $m_h=125\GeV$, the value that we will adopt in the rest of the paper.

Assuming $m_h=125\GeV$, we summarise all data 
in the right panel of Fig.\fig{data} together with their  $1\sigma$ error-bars, as derived by us
following the above-described statistical procedure. 
The horizontal green line in the right panel of Fig.\fig{data} is the SM prediction, and the
horizontal red line is the background-only rate expected in the absence of a Higgs boson.
The grey band shows the $\pm1\sigma$ range for the weighted average of all data.
It lies along the SM prediction.
Furthermore, the global $\chi^2$  of the SM fit is 17 for 15 dof. 

\medskip

However, it is interesting to split data into three categories according to the final states
and compute the average for each one of them:
\beq 
\frac{\hbox{observed rate}}{\hbox{SM rate}} = \left\{\begin{array}{ll}
2.1\pm 0.5 & \hbox{photons} \\
0.5\pm 0.3 & \hbox{vectors: $W$ and $Z$}\\
1.3\pm 0.5 & \hbox{fermions: $b$ and $\tau$}
\end{array}\right.\ .
\label{anomaly}
\eeq
This shows the main anomalous features in current measurements.
First, the $\gamma\gamma$ channels exhibit some excess, mainly driven by the 
vector boson fusion data presented at the Moriond 2012 conference.
Second, there is a  deficit in the vector channels.
Finally, the average rate of fermionic channels  lies along the SM prediction;
here the new Tevatron combination for $h\to b\bar b$ plays
an important r\^{o}le.

If the Higgs boson mass is different from 125 GeV, unlike what indicated by excess in $\gamma\gamma$ distributions
around this value of the invariant mass,
then the $\gamma\gamma$ rate in eq.~(\ref{anomaly}) would be  reduced
with respect to what we assume in our fit.

\section{Reconstructing the Higgs boson properties}
\label{res}

\subsection{Reconstructing the Higgs boson branching fractions}
\begin{figure}[t]
$$\includegraphics[width=0.85\textwidth]{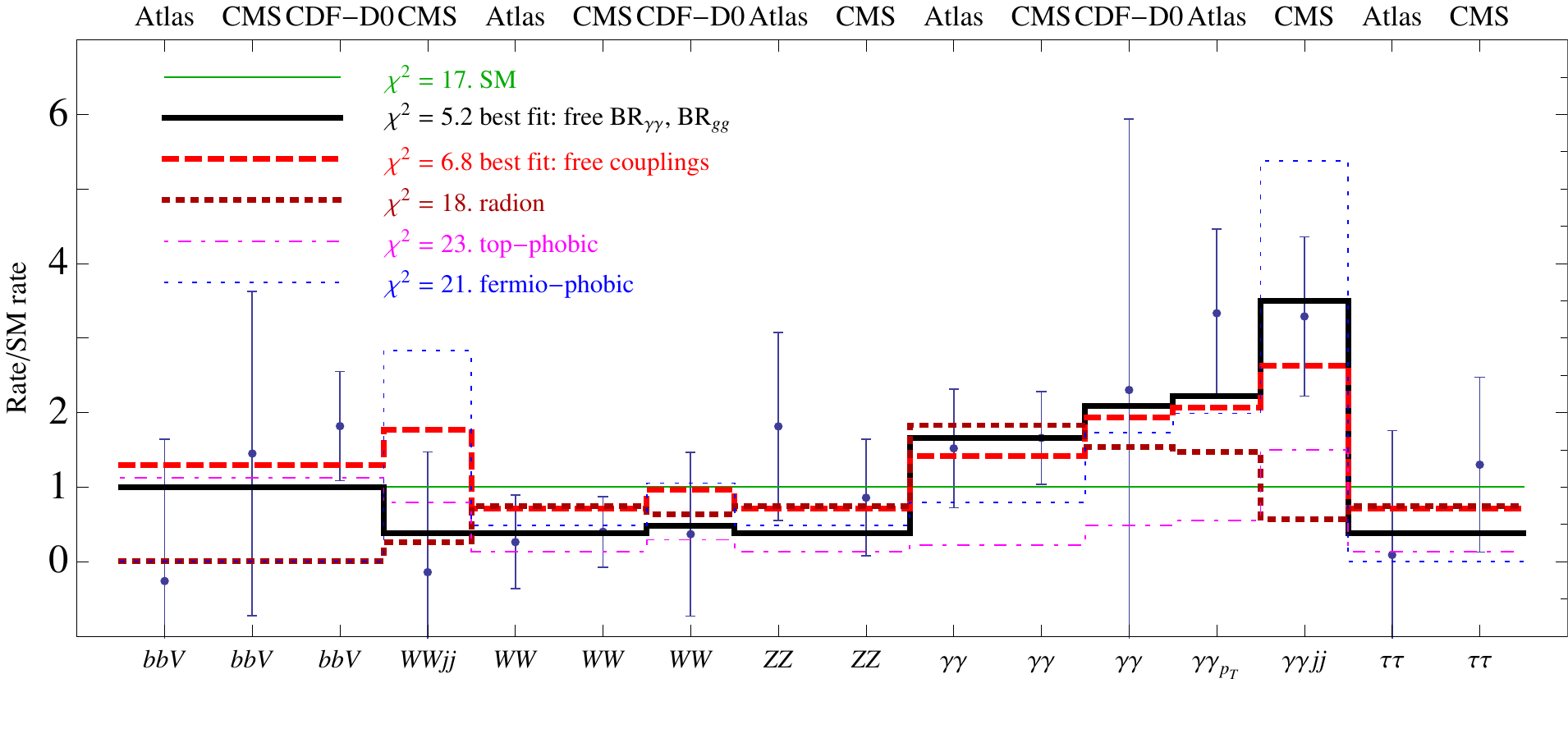}$$
\caption{\em Predictions for the Higgs boson rates in different scenarios: SM, free branching ratios of loop processes,
free couplings, radion, top-phobic and fermiophobic, defined via eqs.~(\ref{allrates},\ref{simplerates},\ref{radionrate}).
\label{fig:data2}}
\end{figure}

The Higgs boson observables that can be most easily affected by new physics contributions 
are those that occur at loop level,  the $h\to \gamma\gamma ,$  $h \to  gg$ and $gg\to h$  rates.
Because the latter two are related via CP, we use a common notation $h\leftrightarrow gg$ to indicate both of them simultaneously. 
Those loop level processes  are particularly relevant for the LHC Higgs boson searches because $\gamma\gamma$ is the cleanest final state, 
and because $gg\to h$ is the dominant Higgs boson production mechanism.
The left panel of Fig.\fig{fitBR} shows, as 
yellow contours with solid borders, the  $ 1\sigma$ and $2\sigma$ ranges  of a global  fit to these two quantities 
 in units of their SM predictions.
The best fit corresponds to
\beq 
\label{BRfit}
\frac{\hbox{BR}(h\leftrightarrow gg)}{\hbox{BR}(h\to gg)_{\rm SM}}\approx 0.3,\qquad
\frac{\hbox{BR}(h\to\gamma\gamma)}{\hbox{BR}(h\to\gamma\gamma)_{\rm SM}}\approx 4,
\eeq
that shows a significant deviation from the SM prediction --- 
the first number allows to best fit the reduced $WW^*$ rates, and the second number allows to fit the enhanced
$\gamma\gamma$ rates,
in agreement with \eq{anomaly}. 
The $\chi^2$ of the global fit is significantly lower with respect to SM, decreasing from 17 (for 15 dof within the SM) to $5.2$ (for 13 dof in this more general fit).
The black thick line in Fig.\fig{data2} shows the best-fit predictions for the various measured rates,
allowing to see how the fit is improved.

The gray region with dotted contours in Fig.\fig{fitBR} show the fit obtained omitting the 
$\gamma\gamma$ rates with cuts dedicated to vector-boson-fusion production (items 6 and 7 in the list above).
In the latter case the agreement with the SM is  improved showing that the such data category plays an important r\^{o}le in the fit.

\subsection{Reconstructing the Higgs boson invisible width}\label{inv}
New physics can easily give a large effect  providing an  extra invisible~\cite{Eboli:2000ze}  Higgs boson decay channel, 
for example into dark matter particles~\cite{Raidal:2011xk,Djouadi:2011aa}.
Alternatively, the effective operator $|\partial_\mu H^\dagger H|^2$ similarly has the effect
of rescaling all rates by a common factor~\cite{Heff}.

In the SM  total Higgs boson width is predicted to be $\Gamma(h)_{\rm SM}\approx 4.0\MeV$ at $m_h=125\GeV$,
too small to be measured directly.

It is well known that measuring the Higgs boson total width at the LHC requires additional assumptions~\cite{hfit}
because the gluon final states cannot be measured over huge QCD background.
Let us explain how present data can probe the Higgs boson width, without directly measuring it.
The gluon fusion production rates are proportional to $\Gamma(gg\to h)$.
In view of approximate CP invariance we can assume that $\Gamma(gg\to h)=\Gamma(h\to gg)$
and we collectively denote them as $h\leftrightarrow gg$.
Then, one partial decay width can be reconstructed by data.
By performing a global fit to the Higgs boson branching ratios in the context of theories
where the decay widths are related 
we can reconstruct the total Higgs boson width.
Of course this is based on theoretical assumptions, 
but the result gets significantly different only in highly deviant models,
e.g.\ in models where the Higgs boson predominantly decays into light quarks (a decay mode not probed by present data).

\begin{figure}[t]
$$\includegraphics[width=0.45\textwidth]{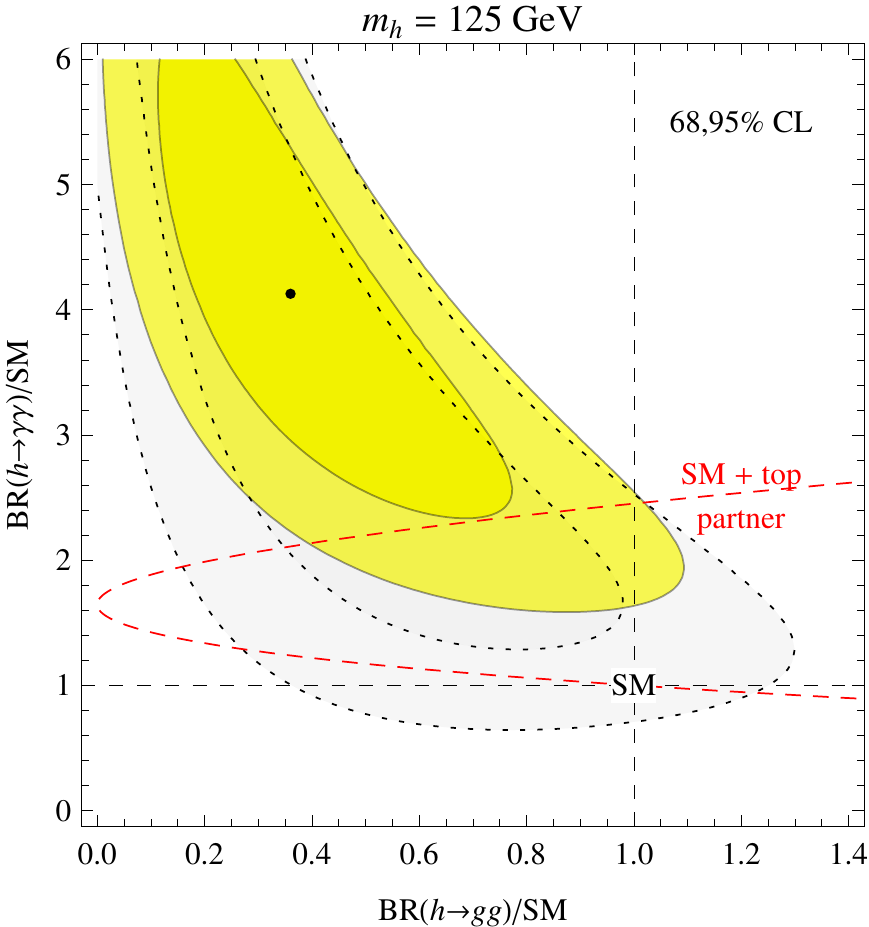}\qquad\includegraphics[width=0.45\textwidth]{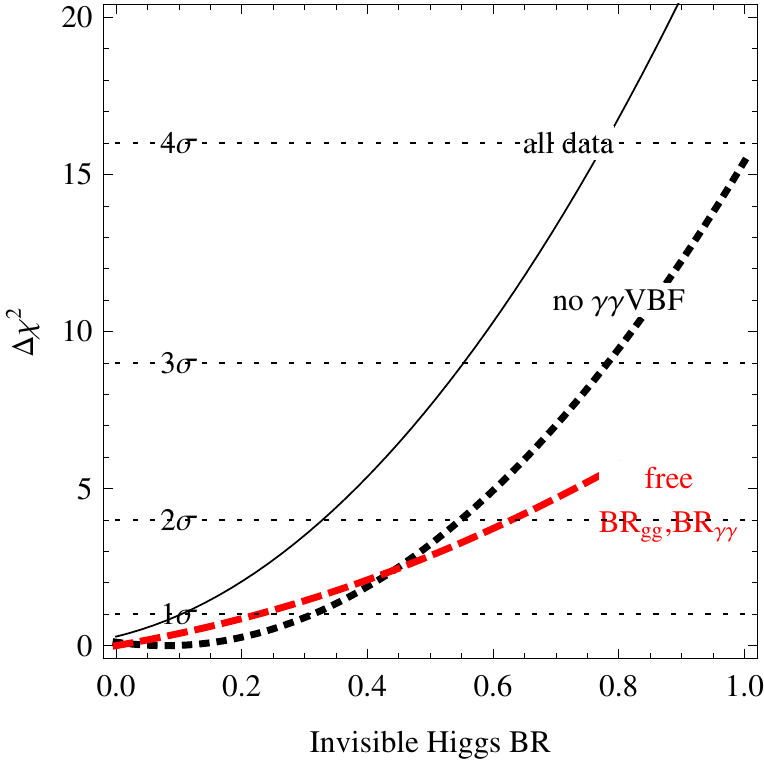}$$ 
\caption{\em {\bf Left}:  fit for the Higgs boson branching fraction to photons and gluons.
In yellow with continuous contour-lines: global fit.
In gray with dotted contour-lines:  the fermiophobic Higgs boson searches are excluded from the data-set.
Red dashed curve: the possible effect of extra top partners, such as the stops.
{\bf Right}:  fits for the invisible Higgs boson branching fraction, under different
model assumptions, as explained in section~\ref{inv}.
\label{fig:fitBR}}
\end{figure}

\medskip

In order to emphasise the mild model-dependence of this fitting procedure we perform three fits under different assumptions.
We show our results ($\chi^2$ as function of the invisible branching ratio) in the right panel of Fig.\fig{fitBR}.
\begin{itemize}
\item[i)] First, we perform a global fit of all data assuming the SM plus an additional invisible decay width,  obtaining
\beq \hbox{BR}(h\to\,{\rm invisible}) = -0.1\pm0.23;\eeq
\item[ii)] Next, we weaken the theoretical assumptions: we
keep the $h\leftrightarrow gg$ and the $h\to\gamma\gamma$ rates as free parameters, and marginalise
with respect to them (red dashed curve).
We see that, even without assuming the SM prediction for $h\leftrightarrow gg$, 
a (weakened)  bound on the Higgs boson invisible width  can still be derived from present data;
the best fit value becomes  positive, but again the preference is not statistically significant.

\item[iii)] Finally, we repeat the fit in i), but ignoring the data for $\gamma\gamma$ from the vector boson fusion channels, obtaining a weaker bound
(dotted curve).
\end{itemize}

Adding an invisible Higgs boson width has the effect of suppressing
all observed rates and, according to our fit, this is not favoured by present data.

\subsection{Reconstructing the Higgs boson couplings}
In this subsection we extract from data the Higgs boson couplings to vectors and fermions, in order to
see if they agree with the SM predictions.
Trying to be as general as possible in describing the Higgs boson couplings,
we proceed phenomenologically extracting from data 
 the following parameters:
\beq  R_W = \frac{g_W}{g_{\rm W}^{\rm SM}},\qquad 
R_Z = \frac{g_Z}{g_{\rm V}^{\rm SM}},\qquad
R_t = \frac{y_t}{y_t^{\rm SM}},\qquad
R_b = \frac{y_b}{y_b^{\rm SM}},\qquad
R_\tau= \frac{y_\tau}{y_\tau^{\rm SM}},
\label{allrates}
\eeq
where $g_W$ is the $WWh$ coupling; $g_Z$ is the $ZZh$ coupling, $y_t$ the top Yukawa coupling,
$y_b$ the bottom Yukawa coupling and $y_\tau$ the tau Yukawa coupling.
All models considered in this work and presented in Fig.\fig{data2} are defined via \eq{allrates}.
The SM corresponds to $R_i=1$ for all the couplings.
These parameter $R_i$ have the following effects:
\begin{itemize}
\item the partonic cross sections for $gg\to h$ and for $gg\to t\bar{t} h$
get rescaled by $R_t^2$;
\item the partonic cross sections for $q\bar q\to q\bar q h$ and for $q\bar{q}\to Vh$ get rescaled by $R_V^2$;
\item the decay widths $h\to VV^*$ get rescaled by $R_V^2$ where $V=\{W,Z\}$;
\item the decay widhts $h\to f\bar{f}$ get rescaled by $R_f^2$ where $f=\{b,\tau,\ldots \}$;
\item the decay width $h\to\gamma\gamma$,
arising from the interference of  one-loop diagrams mediated by the top and by the $W$,
gets rescaled by
$(1.28 R_W - 0.28 R_t)^2$ for $m_h = 125\GeV$;
\item similarly the decay width $h\to Z\gamma$ (not yet measured) gets rescaled by
$(1.05 R_Z - 0.05 R_t)^2$.
\end{itemize}

\begin{figure}[t]
$$\includegraphics[width=0.45\textwidth]{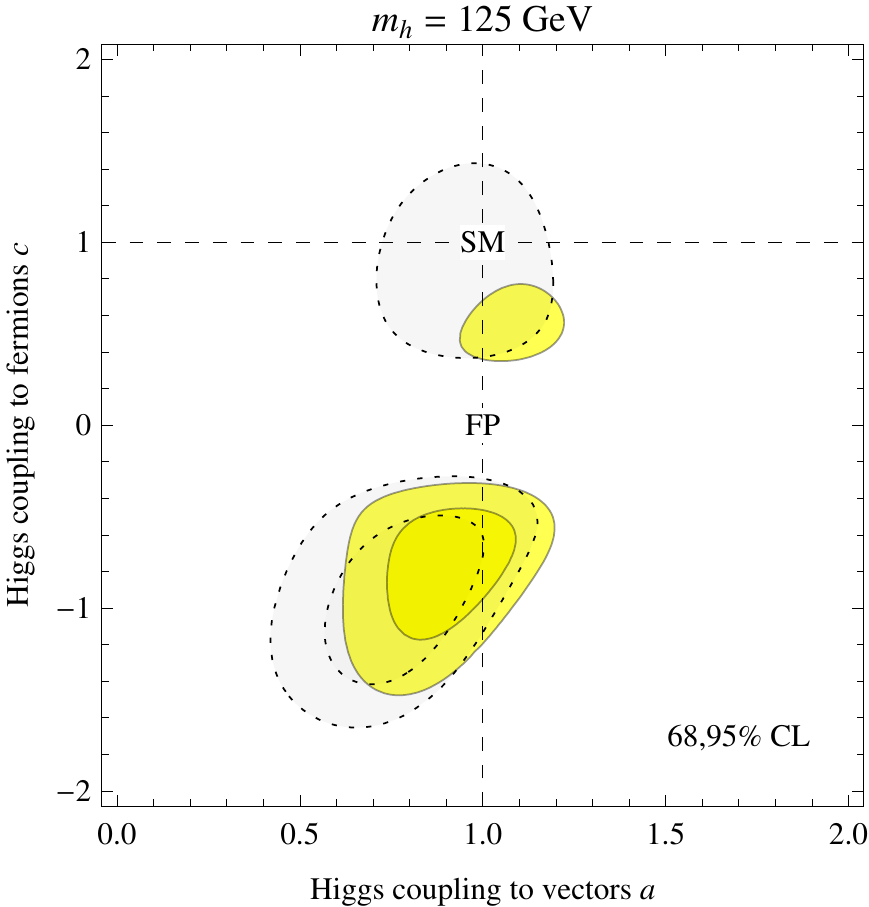}\qquad\includegraphics[width=0.45\textwidth]{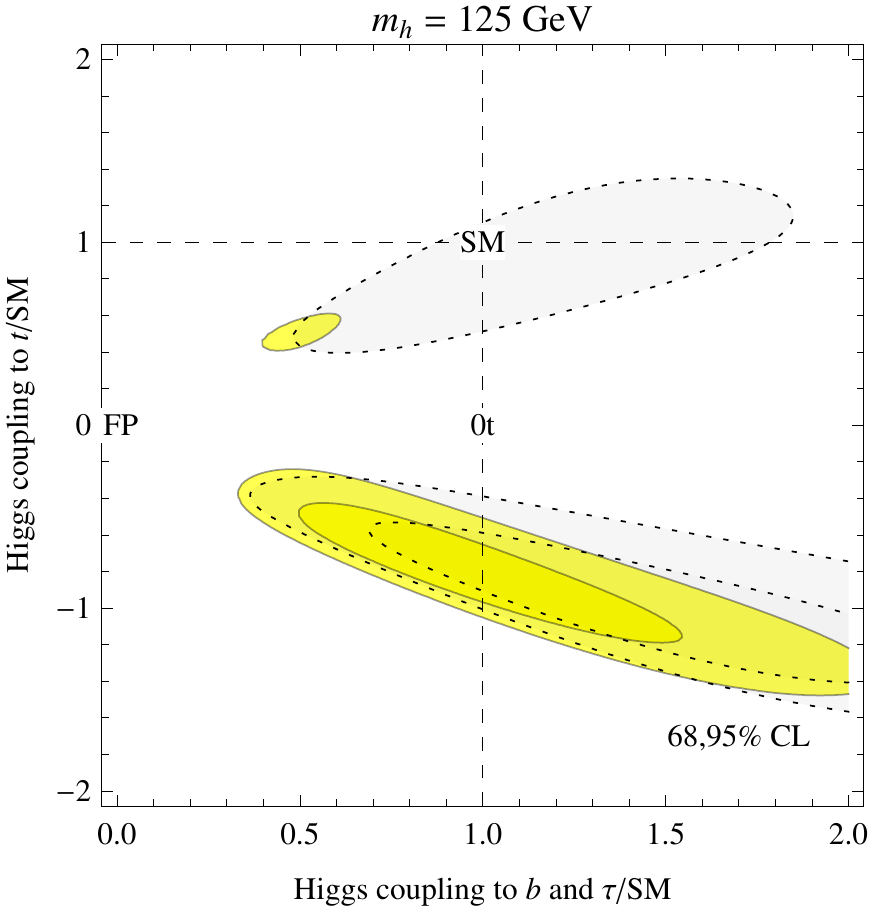}
$$ 
\caption{\em {\bf Left}: fit of the Higgs boson couplings
assuming common rescaling
factors $a$ and $c$ with respect to the SM prediction for vector bosons and fermions,  respectively.  {\bf Right}: fit to the $t$-quark and to $b$-quark and $\tau$-lepton Yukawa couplings assuming the SM couplings to gauge bosons.
The best  fit presently lies somehow away from the SM prediction,
indicated in the figures as `SM'.
The point marked as `FP' is the fermiophobic case, and `0t' denotes the top-phobic case.
Negative values of the top Yukawa coupling are preferred because lead of an enhancement of $h\to \gamma\gamma$.
\label{fig:fitac}}
\end{figure}

\begin{figure}[t]
$$\includegraphics[width=0.48\textwidth]{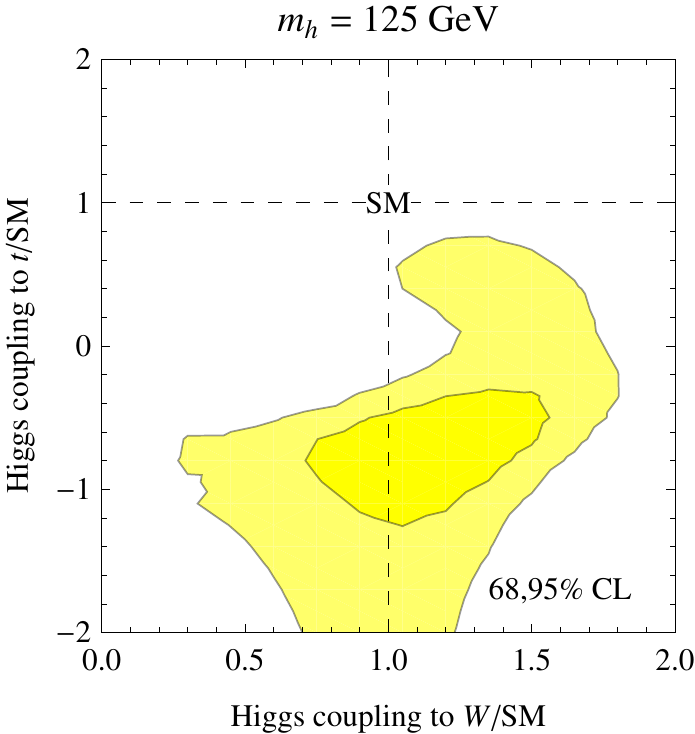}\qquad\includegraphics[width=0.45\textwidth]{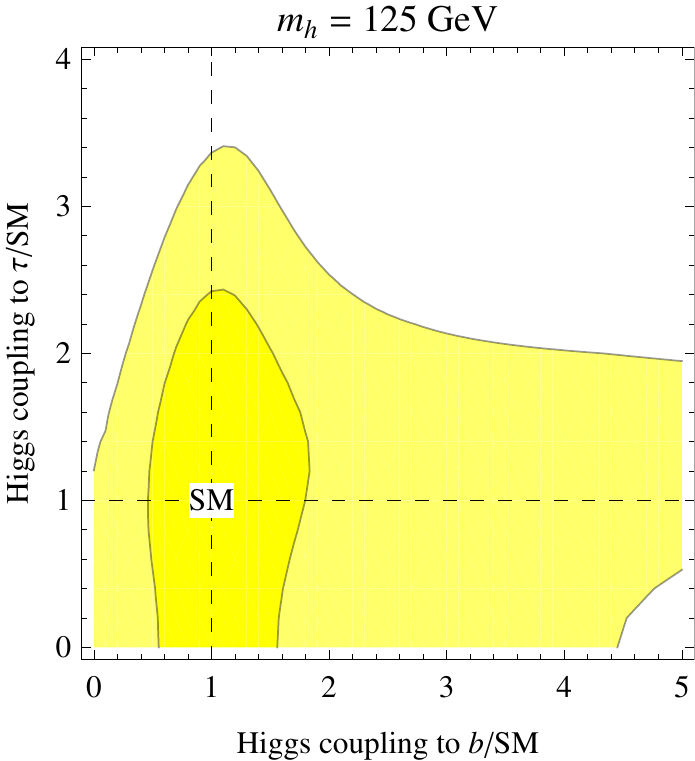}$$ 
\caption{\em Global fit for the Higgs boson couplings to vectors, to the $t$-quark, to the $b$-quark, to the $\tau$ lepton.
All these couplings are freely varied and in each panel we show the $\chi^2$ as function of the  parameters indicated on the axes,
marginalised with respect to all other parameters.  We again assume $m_h=125\GeV$ and find
that the best fit presently lies somehow away from the SM prediction,
indicated in the figures as `SM'.
\label{fig:fitR}}
\end{figure}

A simplifying case considered in previous analyses~\cite{Contino,Falkowski,Grojean}  is a common rescaling factor $a$ for Higgs boson coupling to
vectors and a common rescaling factor $c$ for Higgs boson coupling to fermions:
\beq  
a = R_V\equiv R_W=R_Z,\qquad
c = R_t=R_b=R_\tau.
\label{simplerates}
\eeq
We show in the left panel of Fig.\fig{fitac} the resulting fit (continuous yellow contours).
For comparison the dashed contours show the result obtained ignoring the $\gamma\gamma jj$ data from
CMS and ATLAS, as is also done in Fig.\fig{fitBR}. This allows to compare our results with the ones of previous 
analyses~\cite{Contino,Falkowski,Grojean} (although some other data has also been modified and added by experiments).
Our results essentially agree, up to the difference due to our use of more recent data.

We see that a negative $R_t R_W<0$ is favoured because it implies a constructive interference between the top quark and $W$ boson
loops in the decays $h\to \gamma\gamma$ increasing the corresponding rates. Notice that the new data prefers suppression of 
the $WW^*$ rates via suppression of the $gg\to h$ cross section, while the Higgs boson coupling to vectors can be somewhat larger than
without VBF data. Notice also that the SM point (1,1) is disfavoured beyond $2\sigma.$

%
%
%

%

In the right panel of Fig.\fig{fitac} we assume the SM values for the Higgs boson gauge couplings ($R_W=R_Z=1$) and present a fit to
the Yukawa couplings $R_t$ and $R_b=R_\tau$. We, again, see that $R_t<0$ if somehow favoured and the SM is disfavoured.
The two branches approach the pure fermiophobic point (0,0), denoted by FP in  Fig.\fig{fitac}, but pure fermiophobia is disfavoured by the fit.

In  Fig.\fig{fitR} we consider the most general case where we allow all 4 parameters $R_W=R_Z, R_t,R_b,R_\tau$
to vary and show the favoured regions for the pairs $R_V,R_t$ (left) and $R_b,R_\tau$ (right) marginalised
over the remaining two parameters.
The main features of this global fit remain the same as in previous cases: $R_t R_W<0$ is favoured
and $R_W, R_b$ and $R_\tau$ are constrained to be around their SM values of $1$.
Fig.\fig{data2} shows the best fits (red dashed lines), both allowing for negative Yukawas (thick line) and restricting all Yukawas
to be positive, as in the SM (thin line).

\section{Implications for Higgs boson models}
\label{dis}

In order to  interpret our general results presented in  Figs.\fig{data}-\ref{fig:fitR} in the context of any particular model of
Higgs boson, two logical possibilities arise. 
First, all  the present anomalies in data, listed in  \eq{anomaly},
could be just statistical fluctuations. 
Second, the emerging pattern in \eq{anomaly} could be real and signal new physics beyond the SM 
in the Higgs sector. 
Intermediate possibilities are of course possible.
In order to discriminate between these possibilities, we present in Fig.~\ref{fig:data2} the predictions channel by channel 
of some particular  scenarios that we studied in Figs.\fig{fitBR}-\ref{fig:fitR} for collider searches. 
The best fit $\chi^2$ of those scenarios is also presented in the figure in order to compare different scenarios with each other.

\subsection{The Standard Model}

Naturally, the reference model for all comparisons in the previous sections is  the SM. 
After fixing the Higgs boson mass to the best fit value $m_h=125$~GeV, the SM does not have any free parameter left to vary.
Therefore all the anomalies  in the present data must be statistical fluctuations and disappear with more statistics. 
This interpretation is supported by the fact that  the average of all data agrees  with the SM prediction, 
as seen in Fig.~\ref{fig:data2}, and the global $\chi^2$ is good: 17 for 15 dof
(we recall that with $n\gg1$ degrees of freedom
one expects $\chi^2 = n\pm\sqrt{n}$).

\smallskip

On the other hand, our best fit (black curve in Fig.~\ref{fig:data2}) has a significantly lower
$\chi^2=5.5$ for 13 dof: a bigger reduction than what is typically obtained by adding two extra parameters
(one expects $\Delta \chi^2 =-\Delta n \pm\sqrt{\Delta n}$ when adding $\Delta n\gg1$ parameters).
The SM is disfavoured at more
than $95\%$ CL in this particular context, but of course we added the two parameters that allow to fit the two most apparent anomalies in the data,
the $\gamma\gamma$ excess and the $WW^*$ deficit.

Only more data will tell if this is a trend, or if we are just fitting a statistical fluctuation.

\subsection{Fermiophobia and dysfermiophilia}
Fig.~\ref{fig:data2} shows predictions for different fermiophobic~\cite{Basdevant:1992nb,Gabrielli:2010cw,Gabrielli:2012yz} scenarios.
While bottomphobic Higgs boson is excluded by our fits, top-phobic or pure fermiophobic
Higgs boson (with exactly vanishing Yukawa couplings) provide acceptable fits, of quality
almost as good as the SM fit, despite that their predictions are significantly different.
The pure fermiophobic model captures the features of data qualitatively correctly but predicts larger signal rates
than is observed in the LHC, especially in the $WW^*+jj$ channel.
In addition, the fermiophobic fit suffers from the $h\to b\bar b$ signal claimed by  Tevatron and CMS.

The agreement of the fermiophobic Higgs boson  with data can be improved by
allowing a moderate small additional Higgs boson branching fraction, because this allows to
reduce the too large prediction for the $\gamma\gamma jj$ rate~\cite{Gabrielli:2012yz},
which is very sensitive to the precise value of the Higgs boson mass and width.
In fermiophobic models such small Yukawa couplings can be generated via quantum effects~\cite{Gabrielli:2010cw}.
We note that fermiophobia lowers the vacuum stability bound on the Higgs boson mass, allowing
125~GeV Higgs boson to be consistent with no new physics below Planck scale.

\medskip

Our fits in Figs.\fig{fitac}-\ref{fig:fitR} show that reducing some or all of the SM Yukawa couplings allows to
again significantly improve the global fit compared to the SM, down to $\chi^2 \approx 7$.
The main feature of the improved fit is $y_t\approx -0.7 y_t^{\rm SM}$, because this allows to enhance the
$h\to\gamma\gamma $ rate and reduce the $gg\leftrightarrow h$ rate.
Admittedly, a `wrong' Yukawa coupling to the top and to the other fermions (dysfermiophilia)
is an even more serious pathology than fermiophobia.

\medskip

Various theoretical frameworks easily lead to modified Higgs boson couplings at moderate level.
In models with more than one Higgs multiplet the Yukawa couplings of the light Higgs boson can be non-standard~\cite{Haber:1978jt}
(this is what can happen also in  supersymmetric models).
Alternatively,  in models where
SM fermions mix with extra fermions at the weak scale,
integrating out the extra fermions, their effects get encoded in effective operators 
of the form $\bar{f} f H H^\dagger H$, that lead to modified Higgs boson couplings to the SM fermions $f$~\cite{HHH}.
Such operators also arise in composite Higgs boson models.

In models where the Higgs boson is a composite particle one generically expects that
Higgs boson couplings get modified by form factors, approximated at low energy by effective operators~\cite{Heff}.
This is the framework considered in the fits of Refs.~\cite{Contino,Grojean}.
In this kind of models, the rescaled SM expressions for these rates that we assumed remain valid even 
when new physics is so large~\cite{Vichi}.
The Higgs boson coupling to $W,Z$ vectors can be easily reduced by mixing the Higgs boson with other scalars;
a good fit to electroweak precision data then demands that the extra scalars are not much heavier than the Higgs boson.
Increasing the Higgs boson gauge couplings is theoretically more challenging~\cite{slava}.

\subsection{Supersymmetry}

Supersymmetric theories that attempt to solve the naturalness problem of the electroweak scale
 have been stringently constrained by the LHC direct searches as well as by the Higgs boson results~\cite{MSSM:Higgs,NMSSM:etc}.
Within the MSSM one needs light and strongly mixed stops, and there are 
two main modifications of Higgs physics.
 
\medskip 
 
First, light stops modify the predictions and for the  $h\to\gamma\gamma$ and $gg\leftrightarrow h$ rates \cite{Djouadi:stop}. 
In practice their extra loop effect is described by a deviation of our parameter
$R_t$ from one:
\beq 
R_t = 1+\frac{m_t^2}{4}\left[\frac{1}{m_{\tilde{t}_1}^2}+
\frac{1}{m_{\tilde{t}_2}^2}-\frac{(A_t-\mu/\tan\beta)^2}{m_{\tilde{t}_1}^2m_{\tilde{t}_2}^2}\right],
\eeq
at leading order in the limit $m_{\tilde{t}_{1,2}}\gg m_t$~\cite{Falkowski}.
We see that the sign of the new effect is not fixed and can be negative in the presence of strong stop mixing.

The red dashed curve in the left panel of  Fig.\fig{fitBR} shows how these rates are affected
by $R_t$
(this applies not only to stops, but also to 
any extra particle with same gauge quantum numbers as the top, such as heavy top partners
in little-Higgs models).
$R_t=1$ corresponds to the SM, and $R_t=0$ to the total suppression of $gg\leftrightarrow h$.
A $R_t<1$ increases the $h\to \gamma\gamma$ rate, but only mildly because this rate is dominated by the $W$ loop.
The red dashed curve enters in the best-fit region when $R_t\approx -1.7$, a situation that cannot be
achieved  in view of bounds on the stop mass. 
We recall that such bounds are extremely model-dependent,
because the signature depends on the unknown stop decay modes, and the production depends on the unknown gluino mass.
For example, in gauge-mediated SUSY breaking the lower bound is $310$~GeV \cite{gmsb:stop}. 
If we assume $m_{\tilde{t}_1} \approx m_{\tilde{g}}$,  the bound on the 
stop mass is around $900$~GeV~\cite{gluino:mass}, assuming that gluino decays always via a sbottom
$\tilde{b}$ into $b N_1$ with a neutralino mass $m_{N_1}< 150-300$~GeV. 
The bound on  $m_{\tilde{t}}$ gets about $200$~GeV lower if the gluino decays fully via a stop.

%
\medskip

Second, due to the presence of two Higgs doublets $H_{1}$ and $H_{2}$, one has modification of the Higgs boson couplings at tree level.
Our $R$ parameters get modified as:
\beq R_W =R_Z= \sin (\beta-\alpha),\qquad
R_b = R_\tau = -\frac{\sin\alpha}{\cos\beta},\qquad
R_t = \frac{\cos\alpha}{\sin\beta},\eeq
where $\tan\beta$ is the usual ratio between the two Higgs boson vev, and the $\alpha$ is the usual angle that diagonalises the mass matrix of ${\rm Re}\left( H_{1}^{0},  H_{2}^{0}\right)$, with $\alpha \rightarrow \beta - \pi /2$, in the decoupling limit.
The angles $\alpha$ and $\beta$ depend on the model and
specific deviations arise depending on how $m_h\approx 125\GeV$ is reached:
 extra $D$-terms imply an increase in $h\to b\bar b$ while extra $F$-terms lead to a decrease
 (unless extra singlets are light)~\cite{2Hd}.
 The total $R_t$ is the combination of the two effects discussed above.


As previously discussed, both the
$WW^*$ and the $\gamma\gamma$ rates are roughly proportional to $R_W^2$;
thereby this correlation prevents to go in the direction favoured by data (lower $WW^*$
and higher $\gamma\gamma$), as already observed in the context of numerical MSSM scans~\cite{MSSM:Higgs}, and in extensions of the MSSM \cite{NMSSM:etc}.

\subsection{Dark matter models}

The main motivation for an invisible Higgs boson decay width comes from the existence of Dark Matter (DM) of the Universe.
The Higgs portal \cite{Patt:2006fw} offers a natural possibility to couple the Higgs sector to the dark matter. If the dark matter particles are 
two times lighter than the Higgs boson, they can lead to invisible Higgs boson width. Because Higgs boson decays to fermion
dark matter are essentially ruled out by direct detection constraints~\cite{Djouadi:2011aa}, in this scenario the dark matter is naturally scalar. 

Let us consider, for example, the simplest DM model  obtained adding to the SM a  real singlet scalar field $S$
coupled to the Higgs doublet $H$ by the $-\lambda S^2 |H|^2$ Lagrangian term~\cite{realsinglet}.
Fixing the DM/Higgs boson coupling $\lambda$ assuming that the thermal relic DM abundance is equal to its cosmologically measured value $\Omega_{\rm DM}=0.112 \pm 0.0056$~\cite{WMAP7} allows us to predict the Higgs boson invisible decay width and the direct DM detection cross section $\sigma_{\rm SI}$
\beq \Gamma  (h\to SS) = \frac{\lambda^2V^2}{8\pi m_h}\sqrt{1-4\frac{M_{\rm DM}^2}{m_h^2}},\qquad
\sigma_{\rm SI}
= \frac{\lambda^2 m_N^4 f^2}{\pi M_{\rm DM}^2 m_h^4}
\label{eq:GammaSS}\ .\eeq
The bound $\hbox{BR}_{\rm inv}<0.4$ at 95\% C.L.\ derived in section~\ref{inv} then implies $M_{\rm DM}>50\GeV$
and $\sigma_{\rm SI}<0.4~10^{-44}\,{\rm cm}^2$, assuming the nucleon matrix element $f=0.3$.
While Higgs boson invisible decays to fermionic dark matter are already disfavoured, our work shows that also light scalar dark matter
is not supported by data.

\subsection{Higgs boson or radion?}
The Higgs boson couples to the SM fermions with a strength proportional to fermion masses.
Similar couplings can be obtained by considering an hypothetical particle $\varphi$, a radion,  with a coupling
to the trace of the SM energy-momentum tensor suppressed by some scale $\Lambda$:
\beq \frac{\varphi}{\Lambda} T_\mu^\mu =  \frac{\varphi}{\Lambda} \left(\sum_f m_f \bar f f - M_Z^2 Z_\mu^2 - 2M_W^2 W_\mu^2 + A\right)
\label{radion}
. \eeq
In our language this  is described by setting 
\bea
R\equiv R_W = R_Z = R_t = R_b = R_\tau = \sqrt{2} v/\Lambda,
\label{radionrate}
\eea 
where
$v=174\GeV$.  One important difference arises at quantum level: scale invariance is anomalous,  such that a new term appears,
\beq A = -7\frac{\alpha_3}{8\pi} G_{\mu\nu}^aG_{\mu\nu}^a +\frac{11}{3} \frac{\alpha_{\rm em}}{8\pi} F_{\mu\nu}F_{\mu\nu},\eeq
where the numerical coefficients  are the SM $\beta$-function coefficients for the strong and electromagnetic gauge couplings.
As a result, the $\varphi$ decay widths into $gg$ and $\gamma\gamma$ differ from the corresponding Higgs boson decay widths~\cite{radion}.

Such a particle is often called `radion' because it arises in the context of models with one warped
extra dimension  as the mode that controls its size.
However, in this kind of models the radion can appear
with extra couplings and together with other unseen particles.

We focus on the effective coupling in \eq{radion}, and find that the excess seen around 125 GeV could be due  to such a `radion' rather than to the Higgs boson. The best fit is obtained at $R=0.28\pm0.03$ (i.e.\ $\Lambda \approx  870\GeV$) and its quality is slightly worse that the
best Higgs fit, as illustrated in Fig.\fig{data2}.
More data are needed to discriminate among the two possibilities.


\section{Conclusions}
\label{concl}

We performed a global phenomenological analysis of all Higgs boson collider data available  after Moriond 2012
(including those presented in the context of fermiophobic Higgs boson searches)
assuming that the hints observed at $m_h\approx 125\GeV$ arise from the Higgs boson.
The SM provides an acceptable fit, however it is not favoured: present data with large uncertainties
favor a $h\to\gamma\gamma$ rate enhanced by a factor of $\approx 4$
and a $gg\to h$ rate reduced by a factor of $0.3$.
An invisible Higgs boson branching ratio larger than $0.4$ is disfavoured,
putting constraints on models where dark matter couples to the Higgs boson.
Pure fermiophobic Higgs boson scenario gives almost as good fit as the SM but with significantly different predictions
for the Higgs boson phenomenology. Partially fermiophobic scenarios are among those giving the best global fit.
We find that the apparent excess can alternatively be interpreted as a `radion' i.e.\ a particle similar to the Higgs boson, but
coupled to the trace of the SM energy momentum tensor.

More LHC data should clarify whether the present anomalies in data are statistical fluctuations or first evidence of physics beyond the SM.

\medskip

\paragraph{Acknowledgement} We thank Emidio Gabrielli and  Christophe Grojean for discussions and 
Andrey Korytov and Bill Murray for useful communication.
This work was supported by the ESF grants  8090, 8943, MTT8, MTT60, MJD140 by the recurrent financing SF0690030s09 project
and by  the European Union through the European Regional Development Fund.

\small
\begin{multicols}{2}

\end{multicols}
\end{document}